\documentstyle[mprocl]{article}

\def\be{\begin{equation}}
\def\ee{\end{equation}}
\def\bea{\begin{eqnarray}}
\def\eea{\end{eqnarray}}

\begin{document}

\title{  Can we obtain Conserved  Currents \\ in
 Supersymmetric Quantum 
Cosmology?}

\author{  P.V. Moniz}

\address{DAMTP,  
University of Cambridge\\ Silver Street, Cambridge, CB3 9EW, UK
\\{\small e-mail: {\sf prlvm10@amtp.cam.ac.uk; 
paul.vargasmoniz@ukonline.co.uk}}\\
{\small URL: {\sf http://www.damtp.cam.ac.uk/user/prlvm10}}}

\maketitle\abstracts{
In this paper  we report   whether 
conserved currents 
can be sensibly defined in supersymmetric minisuperspaces. 
Our analysis deals with a  $k=+1$  FRW model. 
Supermatter in the form of scalar supermultiplets 
is included. We 
 show that conserved currents {\em cannot} 
be adequately established except for some very simple scenarios.
More details can be found in ref. \cite{rev,ijmpd}.}

 N=1 supergravity \cite{6} constitutes a   ``square-root'' 
\cite{2}
of
 gravity:  
in finding a physical state $\Psi$,  it is 
sufficient to  solve
the Lorentz and supersymmetry constraints;   
  $ \Psi $ will then consequently  obey 
 the Hamiltonian constraints\footnote{For a review on  
canonical quantization of supersymmetric 
minisuperspaces  see   
ref. \cite{rev}.}. 
Supersymmetry thus 
 induces  advantageous features: 
 in 
many cases we have
to solve   simple {\em first-order}  
differential equations 
in the bosonic variables\cite{rev}.
This   contrasts with the situation without 
supersymmetry:    a 
 {\it second-order}  (Wheeler-DeWitt)  equation
 has to be solved,  employing     boundary conditions 
\cite{5}. 
Therefore, it is quite tempting to address 
from a 
supersymmetric point of view the issue  of 
probability densities for  a 
quantum state $\Psi$ and conservation equations of the type 
 $\nabla_a  J^a = 0$ (see also ref. \cite{ijmpd}).

The approach that we employ here  is based on a differential 
operator representation for the  fermionic variables.
{\em This  constitutes the correct  procedure:}   
it is totally 
 consistent with the existence of second--class constraints  
and subsequent  Dirac brackets in 
supergravity theories. 
These then imply that   fermionic variables and 
their Hermitian conjugates are
 intertwined within  a canonical coordinate--momentum relation.
It should also be pointed out that other 
authors have persued objectives similar to ours 
but with {\em different} methods
\cite{A19,A17a,OOcc}. In particular, 
{\it rigid} 
 supersymmetry 
was used\cite{A19},  
other approach\cite{A17a} was {\it not} 
supersymmetric.  Furthermore,  a  wave function 
arranged as a vector was 
used\cite{OOcc} where  
 the 
Lorentz constraints were severely restricted, but  
an improved approach seems to have been found\cite{OO97}.

Let us consider  
 the action of the more general theory of N=1
 supergravity in the 
presence of gauged supermatter\cite{6}. 
Our physical  variables   include the tetrad 
 $ e^{A A'}_{~~~~\mu} $ 
  and  the   gravitinos which are represented by 
$ 
 \psi^A_{~~\mu}, \bar\psi^{A'}_{~~\mu}$. 
The ``overline'' denotes Hermitian 
conjugation. 
The tetrad for  a $k=+1$ FRW 
model   can be   be written as $
  e_{a\mu} = {\rm diag} \left[
N (\tau), a E_{\hat a i} \right]$,  
where $ \hat a $ and $ i $ run from 1 to 3 and 
$ E_{\hat a i} $ is a basis of left-invariant 1-forms on the unit $ S^3 $
with volume $ \sigma^2 = 2 \pi^2 $.
 We take  
$\psi^A_{~~0} $ and $ \bar\psi^{A'}_{~~0} $ to be  
 functions of time only and $
\psi^A_{~~i} = e^{AA'}_{~~~~i} \bar\psi_{A'}~, ~
\bar\psi^{A'}_{~~i} = e^{AA'}_{~~~~i} \psi_A~.$
We have  introduced  the new spinors $ \psi_A $ and their 
Hermitian  conjugate, $\bar\psi_{A'} $, which
are also functions of time only. 
The scalar supermultiplet 
will consist of   
complex  scalar 
fields $ \phi (t) = r e^{i\theta}, \bar\phi (t)$ 
and their spin-$\frac{1}{2}$ 
partners $ \chi_A (t) , \bar\chi_{A'} (t)$.

Simple Dirac brackets are then 
obtained , namely 
\begin{equation}
 [\chi_{A}, \bar \chi_{B}]_{D} = -i \epsilon_{AB}~, ~
 [\psi_{A}, \bar \psi_{B}]_{D} = i \epsilon_{AB}, 
 [a , \pi_{a}]_{D} = 1~, ~ [\phi, \pi_{\phi}]_{D} = 1~,
 ~[\bar \phi, \pi_{\bar \phi}]_{D} = 1,
\end{equation} 
and the rest of the brackets are zero.
 At this point
 we choose $ (\chi_{A} , \psi_{A} , a , \phi , \bar \phi) $ 
to be the coordinates of the 
configuration space and 
$ (\bar \chi_{A} , \bar \psi_{A}$, 
  $\pi_{a}$ , $\pi_{\phi}$ , $\pi_{\bar \phi}) $ 
to be the momentum operators in this 
representation.
 
The   Lorentz 
constraints take the form 
  $
 J_{AB} = \psi_{(A} \bar{\psi}_{B)} - \chi_{(A} \bar{\chi}_{B)}
 =  0~, $ 
which   implies that  the 
 most general form for the wave function 
of the universe is 
\begin{equation} 
  \Psi = A  +  B \psi^{C} \psi_{C} + C \psi^{C} \chi_{C} + 
D \chi^{C} \chi_{C} + E \psi^{C} \psi_{C} \chi^{D} \chi_{D}, 
\label{eq:2.15}
\end{equation}
where $A$, $B$, $C$, $D$, $E$  are functions 
of $a$, $\phi$ ,$\bar \phi$, 
  only.   
The  bosonic coefficients present in 
eq. (\ref{eq:2.15}) satisfy  attractive relations
\footnote{
Obtained from the supersymmetry constraint equations --- 
$S_A \Psi = 0$ and $\bar S_{A'} \Psi = 0$ --- 
see ref. \cite{rev,ijmpd} for more details.}
in a 3-dimensional minisuperspace:
\begin{eqnarray}
\frac{\partial (A\cdot E)}{\partial a} + 
\frac{\partial (A\cdot E)}{\partial \theta} - ir  \left(
\frac{\partial E}{\partial r} A - \frac{\partial A}{\partial r} E\right) & = &  0,
\label{eq:2.New22a} \\
D_a (B\cdot D)  + 
\frac{\partial (\, B \cdot D)}{\partial \, \theta } 
-ir  \left(\frac{\partial \, B}{\partial \, r} D 
- \frac{\partial \, D}{\partial \, r} B \right)
   & = &   0~,
\label{eq:2.New22b}
\end{eqnarray}
with $D_a = \partial_a - 6/a$. 
However, the presence of 
the terms 
$ ir  \left(
\frac{\partial E}{\partial r} A - \frac{\partial A}{\partial r} E\right)
$ and $
ir  \left(\frac{\partial \, B}{\partial \, r} D 
- \frac{\partial \, D}{\partial \, r} B \right)$ in
 eq. (\ref{eq:2.New22a}) and 
(\ref{eq:2.New22b}), respectively, 
clearly prevent us from obtaining  
 conservation  equations of the type  $\nabla \cdot J = 0$. 
The reason can be identified with  the 
variable $\theta$  no longer being a cyclical 
coordinate when supersymmetry is present (see eq. 
(\ref{eq:2.2pitheta}) below).
To understand this argument, 
let us consider a FRW model with complex scalar fields in 
non-supersymmetric quantum cosmology\cite{Khala}. 
The corresponding action  implies that the 
 conjugate momentum $\pi_\theta \sim   r^2 a^3 
\frac{\partial \, \theta}{\partial \, t}$ 
is a constant and $\theta$ constitutes a cyclical 
coordinate. However, the situation in the corresponding 
supersymmetric scenario is quite different. The 
$\pi_\theta $ takes the form 
\begin{eqnarray}
\pi_\theta & = &
\frac{2\sigma^2}{(1 + r^2)^2} r^2 a^3 
\frac{\partial \, \theta}{\partial \, t} 
+ 
\frac{5 \sigma^2 r^2 a^3}{\sqrt{2}(1 + r^2)^3}
n^{AA'} \bar\chi_{A'}\chi_A -   \frac{3 \sigma^2 r^2 a^3}{\sqrt{2}(1 + r^2)}
n^{AA'} \psi_A \bar\psi_{A'}\nonumber \\
& + & \frac{i r\sigma^2 a^3 e^{-i\theta}}
{\sqrt{2}(1 + r^2)^2}
3 n_{AA'} \chi^A \bar\psi^{A'}
+ \frac{i r \sigma^2 a^3 e^{i\theta}}
{\sqrt{2}(1 + r^2)^2}
3 n_{AA'} \chi^{A'} \psi^{A} \nonumber \\
 & +  & \frac{i r \sigma^2 a^3 e^{-i\theta}}{\sqrt{2}(1 + r^2)^2} 
\chi^A\psi_{0A}  
 - 
\frac{i r \sigma^2 a^3 e^{i\theta}}{\sqrt{2}(1 + r^2)^2} 
 \bar\chi_{A'}\bar\psi_{0}^{A'} ~.
\label{eq:2.2pitheta}
\end{eqnarray}
and thus notice there are
 terms in the 
action that do {\it not} allow 
$\theta$ to be a cyclical coordinate. So, $\pi_\theta$ would not be a 
constant. And this will 
imply the absence of satisfactory
conserved currents.
 A similar situation\footnote{The author is 
grateful to S. Kamenshchik for having pointed out this to him.} 
 would occur in usual quantum cosmology 
with a matter Lagrangian taken from the Wess-Zumino model, 
due to the non-trivial interaction with fermion fields.

Overall, our message in this paper is
that {\em generic} conserved currents do not 
seem feasible  to obtain {\em directly} from the supersymmetry 
constraints equations. 
Only for very simple scenarios does this becomes 
possible. 
Otherwise, conserved currents (and consistent probability 
densities) may  only be    obtained  upon the use 
of 
subsequent Wheeler-DeWitt--like equations. These are 
derived through  the associated 
 supersymmetric algebra of constraints.

In our view, the fundamental reason for our conclusions 
 is related with 
the following.  A physical supersymmetric 
wave functional $\Psi$ 
takes values in a Grassman algebra. Such algebra  
is formed by complex linear combinations of 
products of anti-commuting elements such as 
the gravitino $\psi^A_i$. 
Hence, 
$\Psi \left[e^{AA'}_\mu, \psi^A_\mu, \bar\psi^{A'}_\mu; 
\phi, \bar\phi, \chi^A, \bar\chi^{A'}\right]$ embodies more than a  
  wave function arranged as a vector and satisfying a Dirac-like equation.
Furthermore,  the  first-order differential equations  
derived from the supersymmetry, Lorentzian and Grassmanian-valued 
$\Psi$   constitute  more than a simple 
 set of conditions:  They rather 
represent the action of the supersymmetry constraints 
on {\em different}  fermionic representations of $\Psi$, related by a 
(coordinate--momentum) fermionic Fourier transformation
\cite{rev}.

\section*{Acknowledgements}

\vspace{0.3cm}

This work was supported by   the 
 JNICT/PRAXIS--XXI Fellowship BPD/6095/95. 
The author 
is grateful to 
A. Yu. Kamenshchik
 for useful  comments.
Conversations with   R. Graham, H, Luckock and  O. Obr\'egon 
motivated some of the analysis discussed in this paper and 
ref. \cite{ijmpd}.

\end{document}